# Matrix method for the multi salesmen problem (TSP) with several vehicles

Ishkov Sergey[1], Ishkova Elena[2]

[1] doctor of technical sciences, professor, flight dynamics and control systems, logistics, idpo@ssau.ru,
[2] graduate student, logistics, +7-903-300-1932, ishlena@gmail.com,

Samara state aerospace university (SSAU, Samara, RUSSIA)

Despite the considerable progress made in a transport logistics research, an actual problem today is an optimization of time spends on search for optimal solution [1]. This article describes a method that can be used to simplify and solve the cargo routing problem with several vehicles. The method offers easier and faster solution compare to other known methods. The cargo routing problems are based on the classic traveling salesman problem (TSP). TSP is an NP-hard problem [2]. The task is to find a shortest possible route that visits each travel point exactly once with a return to the starting point given a list of points and their pairwise distances. Modern computer technology is widely used to solve this type of problems. However, practical matters require a more complex undertaking [3]. One of the variations of routing problem is a routing with several vehicles and various cargo limitations [4]. Linear and integer programming methods are often incompatible for this type of problems due to the complexity and vast volume of data processing. Approaches that can be used include an introduction of heuristic methods (hypotheses) [5], [6]. This technique leads to a decomposition of the problem [7].

This article discusses a procedure of division of the routing problem with several vehicles to a problem with one vehicle with matrix approach.

1. Posing the problem showed on Figure 1.

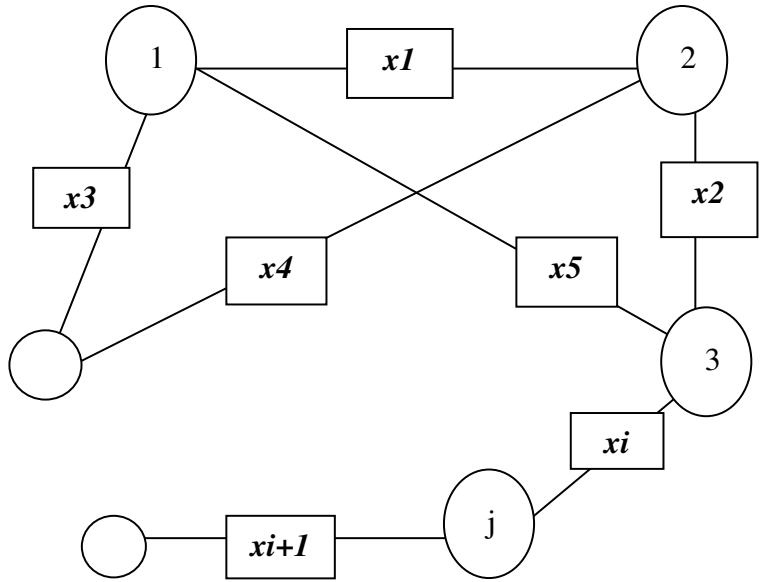

*Figure 1 – Routing problem*

The notation:

$j$ – point, $j \in [1, J]$ ($J$ - the number of points);

$i$ – path, $i \in [1, I]$ ($I$ - the number of paths).

In general, the number of paths (1):

$$I = \frac{J(J-1)}{2} \qquad (1)$$

a) Introduce matrix $\Pi$ (dimension $J{\times}I$) - defining the relationship between the point "$j$" and the path "$i$".

If element $i=0$ to the point "$j$", no visit the point "$j$";

if $i=1$ – visit the point "$j$".

If element $n_{ij} = 0$, the path "$i$" not adjacent the point "$j$";

if $n_{ij} = 1$ - the path "$i$" adjacent the point "$j$".

b) Introduce vector $x_k$ (dimension $I{\times}1$) - defining the route "$k$" vehicle $\forall k = \overline{1, K}$ ("$k$" - the number of used vehicle (vehicles can be similar of different types)). Elements of $x_k$ are Boolean variables.

If element $x_{ki} = 0$ – "$i$" there is no path;

if $x_{ki} = 1$ – "$i$" there is path.

c) Introduce matrix $T_k$ (dimension $1xI$) - defining the cost (time, fuel, cash, etc.) "$k$" vehicle on the movement of "$i$" waterway.

d) Only one vehicle visiting each point. This vehicle returns to the starting point (for example, point $i = 1$) after the transport operation. So, we can say:

$$\Pi \cdot x_k = 2P_k, \quad \forall k = \overline{1, K} \qquad (2)$$

$P_k$ - vector which visiting car # «$k$» all route points (dimension $Jx1$). Elements of $P_k$ are Boolean variables. $P_{kj} \in [0,1]$.

If $P_{kj}=0$ - car # "$k$" didn't visiting a point;

if $P_{kj}=1$ - car # "$k$" visited a point.

If point is visited, it has two adjoin routs – one arriving and one departing.

e) Suppose that each vehicle has two characteristics:
1. $G_k$ – maximum mass cared by vehicle # "$k$", $\forall k = \overline{1, K}$
2. $V_k$ – maximum volume cared by vehicle # "$k$", $\forall k = \overline{1, K}$.

So the restrictions are:

$$\sum_{j=1}^{J} g_j \cdot P_{kj} \le G_k \quad \forall k = \overline{1, K}$$
$$\sum_{j=1}^{J} v_j \cdot P_{kj} \le V_k \quad \forall k = \overline{1, K} \qquad (3),$$

$G_j$, $V_j$ – matrix row (dimension $1xJ$), defining determine the "$g_j$" mass and "$v_j$" volume carried to point "$j$" ($\forall j = \overline{1, J}$).

f) Substituting (2) in (3), we obtain:

$$0{,}5 \cdot g \cdot \Pi \cdot x_k \le G_k \quad \forall k = \overline{1, K}$$
$$0{,}5 \cdot v \cdot \Pi \cdot x_k \le V_k \quad \forall k = \overline{1, K} \qquad (4)$$

g) The objective function is:

$$L = \sum_{k=1}^{K} T_k \cdot x_k \to \min \quad \forall k = \overline{1, K} \qquad (5)$$

To find a shortest possible route for vehicle # "k", it is necessary to define vector $x_k$ for each vehicle by (5) with (4).

This problem is linear programming problem with "KxJ" number of variables and "2K" restrictions.

2. Divide $x_k$ into vectors $x_k^A$ (dimension $J \times 1$) and $x_k^B$ (dimension $(I-J) \times 1$).

$$x_k = [x_k^A, x_k^B] \qquad (6)$$

Similarly, we divide $T_K$:

$$T_k = [T_k^A; T_k^B]$$

$$\Pi = [\Pi^A; \Pi^B]$$

$\Pi^A$ (dimension $J \times J$) – square matrix, $\Pi^B$ – matrix with dimension $J \times (I-J)$.

The divisions cannot be done arbitrarily, because the resulting matrix $\Pi^A$ should not be singular, i.e. have an inverse $(\Pi^A)^{-1}$.

(2) with (6):

$$2P_k = (\Pi^A \cdot x_k^A + \Pi^B \cdot x_k^B)$$

$x_k^A$ equally:

$$x_k^A = 2(\Pi^A)^{-1} \cdot P_k - (\Pi^A)^{-1} \cdot \Pi^B \cdot x_k^B \qquad (7)$$

So, the objective function with (7) is:

$$L = \sum_{k=1}^{K} \left\{ T_k^A \cdot 2\left[(\Pi^A)^{-1} \cdot P_k - (\Pi^A)^{-1} \cdot \Pi^B \cdot x_k^B\right] + T_k^B \cdot x_k^B \right\} \to \min \qquad (8)$$

$$L = 2\sum_{k=1}^{K} T_k^A \cdot (\Pi^A)^{-1} \cdot P_k + \sum_{k=1}^{K} \left[-T_k^A \cdot (\Pi^A)^{-1} \cdot \Pi^B \cdot x_k^B + T_k^B \cdot x_k^B\right] \to \min$$

Introduce:

$$M_k = 2T_k^A (\Pi^A)^{-1}$$

$$L^* = \sum_{k=1}^{K} M_k \cdot P_k \qquad (9)$$

$$L_0 = \sum_{k=1}^{K} \left\{-T_k^A \cdot (\Pi^A)^{-1} \cdot \Pi^B \cdot x_k^B + T_k^B \cdot x_k^B\right\} \to \min$$

So objective function divides into two parts: one part depends on $P_k$; another part depends on $x_k^B$. (3) and (4) depend only on $P_k$. Therefore, problem is divided into two independent problems.

$$L = L^* + L_0 \to \min \tag{10}$$

$$\begin{aligned} g \cdot P_k &\leq G_k \\ v \cdot P_k &\leq V_k \end{aligned} \tag{11}$$

$$\sum_{k=1}^{K} P_{jk} = 1 \quad \forall j = 2, J \tag{12}$$

$$P_{1k} = 1 \quad \forall k = 1, K \tag{13}$$

Equation (12) means that only one vehicle visit every point, except first.

Therefore, in accordance with (10), the general optimization problem can be separated into a problem of determining $P_k$ for each vehicle with (11) and (13) followed by the classical problem of routing for each vehicle solution.

Inversion $\Pi^A$ is the main computational problem. Inversion can impose significant restrictions on the matrix method for a large size problem.

We formed the following rule of formation of the $\Pi^A$ matrix, allowing its reversibility:

If an <u>odd</u> number of points vectors $x^A$ visits will include only those paths which form a closed path with the return to the starting point $i = 1$;

If an <u>even</u> number of points vectors $x^A$ visit will include paths of forming opened route.

Analytical formulas for calculating the matrix $M_k$ for arbitrary $J$:

$$M_{k, j, i} = (-1)^j \left[ -\sum_{i=1}^{j-1}(-1)^i T_i + \sum_{i=j}^{k}(-1)^i T_i \right] \tag{14}$$

(14) divides the problem without resorting matrix inversion $\Pi^A$.

3. Solve the routing problem with eleven points and the three vehicles with matrix method.

Find a shortest possible route that visits each point exactly once with a return to the starting point. Find $P^*_k$ ($k = 3$). The objective function is minimal.

Graph distance matrices (Table 5). Cell count is a number of paths ($X_{kj}$). Row and column headers - number of points to which the adjacent road $X_{kj}$.

Table 5 - Number of routes between the eleventh points

|    | 1  | 2  | 3  | 4  | 5  | 6  | 7  | 8  | 9  | 10 | 11 |
|----|----|----|----|----|----|----|----|----|----|----|----|
| 1  | XX | 2  | 12 | 13 | 14 | 15 | 16 | 17 | 18 | 19 | 1  |
| 2  | 2  | XX | 3  | 28 | 29 | 31 | 34 | 35 | 42 | 49 | 20 |
| 3  | 12 | 3  | XX | 4  | 30 | 32 | 41 | 36 | 50 | 43 | 21 |
| 4  | 13 | 28 | 4  | XX | 5  | 33 | 40 | 46 | 37 | 51 | 22 |
| 5  | 14 | 29 | 30 | 5  | XX | 6  | 39 | 47 | 52 | 38 | 23 |
| 6  | 15 | 31 | 32 | 33 | 6  | XX | 7  | 48 | 44 | 53 | 24 |
| 7  | 16 | 34 | 41 | 40 | 39 | 7  | XX | 8  | 54 | 45 | 25 |
| 8  | 17 | 35 | 36 | 46 | 47 | 48 | 8  | XX | 9  | 55 | 26 |
| 9  | 18 | 42 | 50 | 37 | 52 | 44 | 54 | 9  | XX | 10 | 27 |
| 10 | 19 | 49 | 43 | 51 | 38 | 53 | 45 | 55 | 10 | XX | 11 |
| 11 | 1  | 20 | 21 | 22 | 23 | 24 | 25 | 26 | 27 | 11 | XX |

In accordance with table 5 the total number of points is $J = 11$ (first point is base). The total number of paths is $I = 55$.

The cost of moving each vehicle $T^*_k$ ($k = 3$) on the relevant routes is

$T^*_1$ = [7 6 8 11 8 10 9 4 5 8 6 3 4 5 6 4 3 5 6 5 4 3 6 6 5 4 3 7 6 5 4 8 7 6 5 10 8 7 6 12 10 8 7 6 5 4 3 10 12 7 8 8 6 5 12]

$T^*_2 = 1.2 \cdot T_1$,   $T^*_3 = 1.25 \cdot T_1$.

The main problem of this type of problem solving matrix method is the problem of finding the inverse matrix $(\Pi^A)^{-1}$. How we known:

$\Pi = [\Pi^A, \Pi^B]$,

$\Pi$ - defining the relationship between the point "$j$" and the path "$i$". So, elements of $\Pi_{ji}$ can take two values (1 and 0).

If $\Pi_{ji} = 0$ - the path # "$i$" connect with point # "$j$";

if $\Pi_{ji} = 1$ - the path # "$i$" didn't connect with point # "$j$".

So, in accordance with the rule:

$T_1^A = [7\ 6\ 8\ 11\ 8\ 10\ 9\ 4\ 5\ 8\ 6]$

$T_1^B = [3\ 4\ 5\ 6\ 4\ 3\ 5\ 6\ 5\ 4\ 3\ 6\ 6\ 5\ 4\ 3\ 7\ 6\ 5\ 4\ 8\ 7\ 6$
$5\ 10\ 8\ 7\ 6\ 12\ 10\ 8\ 7\ 6\ 5\ 4\ 3\ 10\ 12\ 7\ 8\ 8\ 6\ 5\ 12]$

$T_2^A = 1.2 \cdot T_1^A,\ T_2^B = 1.2 \cdot T_1^B$

$T_3^A = 1.25 \cdot T_1^A,\ T_3^B = 1.25 \cdot T_1^B$

We will solve this problem a few deliveries: with and without taking into account constraints on the mass of the cargo.

*3.1 Solution of the problem without constraints*

Find a shortest possible route that visits each point exactly once with a return to the starting point. Find $P^*_k$ ($k = 3$). The objective function is minimal.

$$L^* = M^*_1 P^*_1 + M^*_2 P^*_2 + M^*_3 P^*_3 \to min \qquad (15)$$

Moreover, only one vehicle can visit a fined point.

$L^* = 11P_{1\,2} + P_{1\,3} + 14P_{1\,4} + 8P_{1\,5} + 8P_{1\,6} + 12P_{1\,7} + 5P_{1\,8} + 3P_{1\,9} + 6P_{1\,10} + 9P_{1\,11} +$

$+ 1.2\,(11P_{2\,2} + P_{2\,3} + 14P_{2\,4} + 8P_{2\,5} + 8P_{2\,6} + 12P_{2\,7} + 5P_{2\,8} + 3P_{2\,9} + 6P_{2\,10} + 9P_{2\,11}) + \qquad (16)$

$+ 1.25\,(11P_{3\,2} + P_{3\,3} + 14P_{3\,4} + 8P_{3\,5} + 8P_{3\,6} + 12P_{3\,7} + 5P_{3\,8} + 3P_{3\,9} + 6P_{3\,10} + 9P_{3\,11}) \to min$

Solving each problem of each vehicle with branch and bound method. Optimal routes with minimum of objective function are:

First vehicle route is 1-11-9-7-10-6-2-4-8-5-3-1.

Second and Third vehicles are not used for the transport operation. The value of the objective function is L = 46.

*3.2 The problem with the restrictions on the mass carried by each vehicle*

Capacity vehicles are $G_k$ ($k = 3$):

$G_1 = 3$ – first vehicle capacity;

$G_2 = 5$ – second vehicle capacity;

$G_3 = 15$ – third vehicle capacity;

Matrix distribution of goods on the points is $G^* = [0\ 2\ 1\ 3\ 1\ 2\ 2\ 3\ 4\ 0.5\ 0.5]$.

Find a shortest possible route that visits each point exactly once with a return to the starting point. Find $P^*_k$ ($k = 3$). The objective function (16) is minimal. Moreover, only one vehicle can visit a fined point.

$$2P_{1\,2} + P_{1\,3} + 3P_{1\,4} + P_{1\,5} + 2P_{1\,6} + 2P_{1\,7} + 3P_{1\,8} + 4P_{1\,9} + 0{,}5P_{1\,10} + 0{,}5P_{1\,11} \leq 3$$
$$2P_{2\,2} + P_{2\,3} + 3P_{2\,4} + P_{2\,5} + 2P_{2\,6} + 2P_{2\,7} + 3P_{2\,8} + 4P_{2\,9} + 0{,}5P_{2\,10} + 0{,}5P_{2\,11} \leq 5 \quad (17)$$
$$2P_{3\,2} + P_{3\,3} + 3P_{3\,4} + P_{3\,5} + 2P_{3\,6} + 2P_{3\,7} + 3P_{3\,8} + 4P_{3\,9} + 0{,}5P_{3\,10} + 0{,}5P_{3\,11} \leq 15$$

So, the resulting integer linear programming problem is solution of (16), (17) and find $P^*_1$, $P^*_2$, $P^*_3$.

$P^*_1 = [0\ 0\ 0\ 0\ 0\ 1\ 0\ 0\ 1\ 1]$ - First vehicle visiting point # 1 (base), 7, 10, 11.

$P^*_2 = [1\ 0\ 0\ 1\ 1\ 0\ 0\ 0\ 0\ 0]$ - Second vehicle visiting point # 1 (base), 2, 5, 6.

$P^*_3 = [0\ 1\ 1\ 0\ 0\ 0\ 1\ 1\ 0\ 0]$ - Third vehicle visiting point # 1 (base), 3, 4, 8, 9.

Solving each problem of each vehicle with branch and bound method. Optimal routes with minimum of objective function are:

First vehicle route is 1-7-10-11-1;

Second vehicle route is 1-5-2-6-1;

Third vehicle route is 1-3-9-8-4-1.

The value of the objective function is L = 63.95.

*3.3 The problem with the restrictions on the mass and volume carried by each vehicle*

Capacity vehicles are $G_k$ ($k = 3$):

$G_1 = 10$ – first vehicle capacity;

$G_2 = 15$ – second vehicle capacity;

$G_3 = 18$ – third vehicle capacity.

Tonnage vehicles are $V_k$ ($k=3$):

$V_1 = 15$ – first vehicle tonnage;

$V_2 = 20$ – second vehicle tonnage;

$V_3 = 40$ – third vehicle tonnage.

Matrix distribution of goods on the points is $G^* = [0\ 2\ 1\ 3\ 1\ 2\ 2\ 3\ 4\ 0.5\ 0.5]$, $V^* = [0\ 10\ 12\ 13\ 11\ 2\ 2\ 3\ 4\ 1\ 3]$.

Find a shortest possible route that visits each point exactly once with a return to the starting point. Find $P^*_k$ ($k = 3$). The objective function (16) is minimal. Moreover, only one vehicle can visit a fined point.

$$2P_{12} + P_{13} + 3P_{14} + P_{15} + 2P_{16} + 2P_{17} + 3P_{18} + 4P_{19} + 0{,}5P_{1\,10} + 0{,}5P_{1\,11} \leq 10$$
$$2P_{22} + P_{23} + 3P_{24} + P_{25} + 2P_{26} + 2P_{27} + 3P_{28} + 4P_{29} + 0{,}5P_{2\,10} + 0{,}5P_{2\,11} \leq 15 \quad (18)$$
$$2P_{32} + P_{33} + 3P_{34} + P_{35} + 2P_{36} + 2P_{37} + 3P_{38} + 4P_{39} + 0{,}5P_{3\,10} + 0{,}5P_{3\,11} \leq 18$$

$$10P_{12} + 12P_{13} + 13P_{14} + 11P_{15} + 2P_{16} + 2P_{17} + 3P_{18} + 4P_{19} + 1P_{1\,10} + 3P_{1\,11} \leq 15$$
$$10P_{22} + 12P_{23} + 13P_{24} + 11P_{25} + 2P_{26} + 2P_{27} + 3P_{28} + 4P_{29} + 1P_{2\,10} + 3P_{2\,11} \leq 20 \quad (19)$$
$$10P_{32} + 12P_{33} + 13P_{34} + 11P_{35} + 2P_{36} + 2P_{37} + 3P_{38} + 4P_{39} + 1P_{3\,10} + 3P_{3\,11} \leq 40$$

So, the resulting integer linear programming problem is solution of (16), (18), (19) and find $P^*_1$, $P^*_2$, $P^*_3$.

$P^*_1 = [0\ 0\ 0\ 0\ 1\ 1\ 1\ 0\ 1\ 1]$ - First vehicle visiting point # 1 (base), 6, 7, 8, 10, 11.

$P^*_2 = [1\ 0\ 0\ 0\ 0\ 0\ 1\ 0\ 0]$ - Second vehicle visiting point # 1 (base), 2, 9.

$P^*_3 = [0\ 1\ 1\ 1\ 0\ 0\ 0\ 0\ 0]$ - Third vehicle visiting point # 1 (base), 3, 4, 5.

Solving each problem of each vehicle with branch and bound method. Optimal routes with minimum of objective function are:

First vehicle route is 1-11-6-10-7-8-1;

Second vehicle route is 1-9-2-1;

Third vehicle route is 1-3-5-4-1.

The value of the objective function is L = 72.8.

Therefore, matrix approaches division of the routing problem with several vehicles to a problem with one vehicle. The problem with 55 variables and more than 40 restrictions divided into some different independent problems with less size and time of solve. This method offers easier and faster solution compare to other known methods.